# Study of nonlinear optical diffraction patterns using machine learning models based on ResNet 152 architecture

BEHNAM PISHNAMAZI,[1] EHSAN KOUSHKI,[2,*]

[1]Undergraduate student, physics Department, Hakim Sabzevari University, razavi khorasan, Iran
[2]Assitant professor, physics Department, Hakim Sabzevari University, razavi khorasan, Iran

**Abstract:** As the advancements in the field of artificial intelligence and nonlinear optics continues new methods can be used to better describe and determine nonlinear optical phenomena. In this research we aimed to analyze the diffraction patterns of an organic material and determine the nonlinear optical phase change and nonlinear refraction index of the material in question by utilizing ResNet 152 convolutional neural network architecture in the regions of laser intensity that the diffraction rings are not clearly distinguishable. This approach can open new sights for optical material characterization in situations where the conventional methods do not apply.

## 1. Introduction

In the last decades nonlinear optical materials attracted a lot of attention due to their wide applications in optical modulators [1], eye protectors [2], optical switchers [3] and mode-locking [4]. With the recent advances in the field of nonlinear optics, the need for new and more accurate techniques for measuring nonlinear optical properties of materials is inevitable. For instance one of the main problems in measuring the nonlinear refractive index of Kerr type third order nonlinear materials is the fact that for high optical nonlinearities, side effects of phase change disturb the process of measurement. In Z-scan measurements [5] [6] [7] nonlinear refractive index is measured using the closed aperture setup in which the passed beam thorough the nonlinear sample is confined using a limited circular aperture. Hence the passing power through the aperture is sensitive to divergence or convergence of the beam, therefore nonlinear refractive index can be measured using transmitted power signals [6] [8]. In spite of this in high nonlinearities diffraction rings pattern occurs at the apertures plane which causes distortions in the passing power and disables close aperture Z-scan measurements. In these cases if the diffraction rings are distinguishable and countable the value of the nonlinear refractive index can be easily obtained [9]. In cases which the rings are not easily distinguishable due to various reasons such as, nonlinear phase change is not enough, local thermal convection currents in aqua samples or diffraction, an alternative method such as machine learning can be extremely useful.

Image classification as a branch of supervised machine learning tries to distinguish between two (Binary) or several types of images (Categorical) by analyzing the images and discovering patterns that allows it to successfully determine the type or label of the image. Convolutional Neural networks (CNN) are the most common models that we use to create a network that best achieves this goal [10]. One of the key elements in creating a suitable CNN is the architecture of the neural network. In this study we used ResNet 152 Convolutional Neural network architecture and fine-tuned it to best serve our purpose of first classifying and mediation of optical intensities from diffraction rings patterns and second, to help us determine the nonlinear optical phase change and nonlinear refraction index 'n₂' in materials and situations where the diffraction ring patterns are not easily distinguishable [11] and cannot be directly derived from conventional theoretical calculations.

## 2. Materials and method

### 2.1. Materials

For this study we needed a relatively strong nonlinear material, since we needed a material that produces blurry diffraction rings in order for us to be able to use machine learning techniques to count precisely the number of rings present in different intensities and calculate the nonlinear phase change of the material in question. After many considerations we opted to use a $37.5 \times 10^{-4}$ gr/ml solution of 2, 6-Diamino-4-(3, 4-dihydroxyphenyl) pyridine-3, 5-dicarbonitrile (4) as our primary material (Fig.1) and DMSO as our solvent [12].

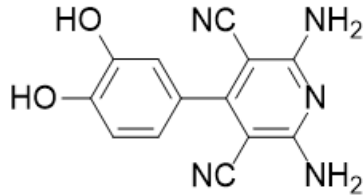

Fig.1 Molecular structure of the compound.

### 2.2. Experimental setup

We performed this experiment on the Z-scan setup since this technique is very effective for achieving good resolution images when we are dealing with diffraction rings [13]. As seen in (Fig.2) our 533 nm green laser beam passes thought a lens so that the laser beam can be focused inside the cell that contains the solution, after that another lens will create a secondary focal point so that we can control where the image is formed, end product of our setup is a clear focused image of the diffraction rings pattern. By passing the beam through the sample, it experiences an optical phase change due to its nonlinearity:

$$\Delta \varphi_0 = \frac{2\pi}{\lambda} (n_2 I) L_{ef} \qquad (1)$$

where λ, I and $L_{ef}$ are the wavelength of the laser, intensity of the laser at the focal point and the effective length of the sample, respectively. In cases that the rings patterns occur, there is a clear relation between the number of rings (N) and the optical phase change:

$$N = \frac{\Delta \varphi_0}{2\pi} \qquad (2)$$

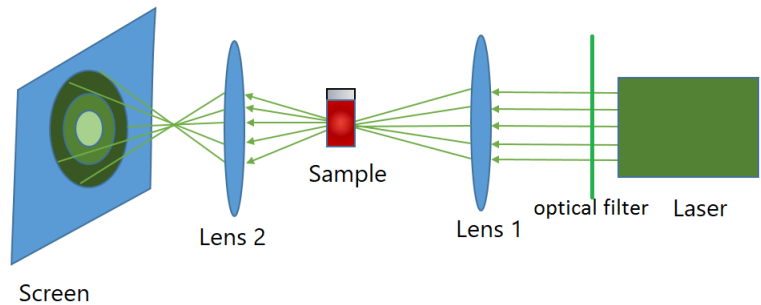
Fig.2. Schematic description of the optical setup.

*2.3. Image preprocessing and pattern extraction*

Almost all image classification models use some level of image preprocessing[14], the most common form of it being dividing the image pixel values by 255 to get values between zero and one[15]. Thanks to Tensorflow we were not required to do all the preprocessing for the ResNet architecture by hand and by using the proper generator we were able to do all the preprocessing needed for our model with few lines of code.

For our model we took the raw image obtained directly from the experiment setup and by using the generator resized it to an (224, 224, 3) image, we also done operations on pixel values and set them to range from -1 to 1, moreover our RGB values were not switched during this preprocessing (unlike VGG16 that switches red and blue) [16] and we did not felt the need for using data augmentation as the spherical symmetry of our experiments prevent any meaningful improvements [17].

Later we split our data into training and validation sets, %85 of our data for training set and %15 for validation. Testing set data wasn't meant to be in any of these groups so that our network would be tested against images that it had never seen before.

Our model starts with identifying interest point in the image then it uses the number of rings and their thickness to better understand and form a relation between the label which is the intensity and the image that it observes. Our model takes into account several more futures in the image such as particles that appear after a certain intensity threshold or the formation and increasing brightness of rings at higher intensities.

*2.4. Neural network architecture*

One of the key element in creating a suitable Convolutional Neural networks (CNN) is the architecture of the neural network. In this study we used ResNet 152 Convolutional Neural network architecture and modified it to best serve our purpose of classifying optical intensities and number of rings present in different intensities [18]. First we needed to remove the last Dense layer of the ResNet 152 so that we can replace it with our own Artificial neural network (ANN).our ANN consists a flatten layer and 3 Dense layers that each one has respectively 2048,60,4 nodes or neurons (Fig.3) [19]. We used stochastic gradient decent (SGD) as our optimizer and settled for 0.0001 for our learning rate [20]. Also we used 'softmax' as our activation function for the prediction layer and 'relu' for the rest of our ANN layers, in the end we used 'CategoricalCrossentropy 'as our loss function [21]. After sufficient preprocessing for the ResNet architecture our model receives images from a generator in (224, 224, 3) format. After that our model starts to learn patterns inside the images and begins to update its weights.

Maybe the most important advantage of ResNet on other image classification models is the presence of identity shortcut connections that allows ResNet to better compensate with the vanishing gradient issue [22].

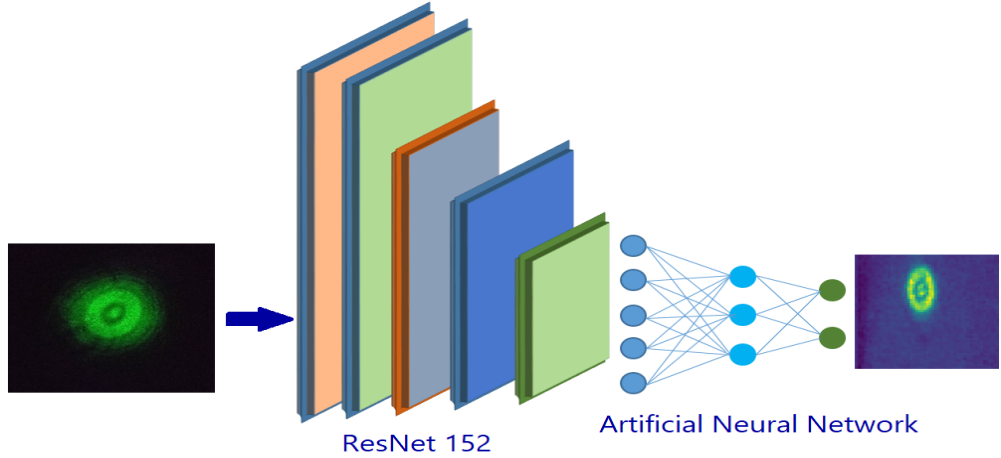

Fig.3 Schematic of the network's Architecture

## 3. Results and discussion

### 3.1 Training & validation sets Results

As training begins our model slowly starts to learn patterns and key futures in the images with intensities ranging from 145.15 w/cm$^2$ to 1314.22 w/cm$^2$ and starts to update its weight and minimize its loss function value. (Fig4.A)

Our model learns at a very stable rate and performs with an accuracy close to 98% on the training set. A steady rise in AUC score shows that our results are actually reliable and we do not suffer from false positive outcomes, at the end we achieved 0.9921 AUC score. (Fig4.B)

To avoid the illusion of learning in our model, we gave our trained model a set of images that it hadn't trained upon in each epoch. This action allows us to verify if our model is truly learning key patterns in the images given to it or is it just memorizing labels with a set of images that it hasn't seen in the training set. If our model is memorizing it would perform poorly in the validation set, but our model predicts the validation set by approximately 96% accuracy which shows that our model is actually learning and can generalize really good on images that it was not originally trained on. Once again we see that our model provides reliable results, an AUC score of 0.9967 on validation set is what our model was able to produce. (Fig4.C)

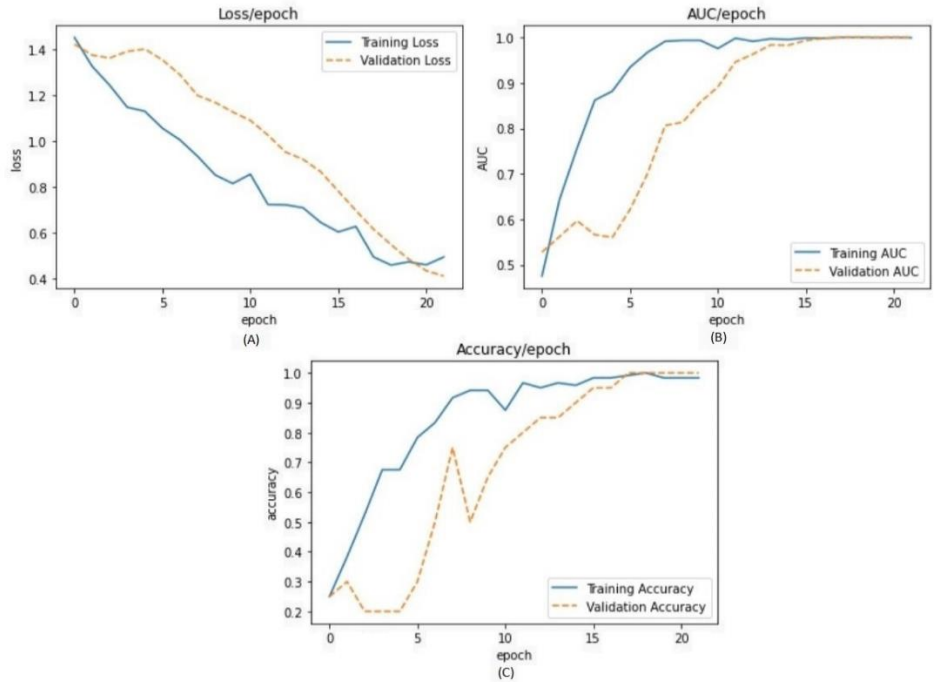

Fig.4 Graphs of the model's loss, AUC and Accuracy, respectively

### 3.2 Testing set approach & results

Our approach for testing set wasn't to select intensities that were already present in our training or validation sets, instead we opted for intensities outside the range of training and validation sets so that we test how good our model can detect intensities that it has never seen or even learned upon before. This approach rises from the fact that as we go higher in intensity the diffraction patterns that we observe do not change as dramatically as in lower intensities, this is a result of a phenomenon known as optical phase saturation [23]. This phenomenon allow us to be able to narrow our expected intensity up to a certain range and will be beneficial for us in terms of saving time, equipment and labor. Our model successfully classified 1022.64 w/cm$^2$ intensity as 1314.22 w/cm$^2$ with 97.5% accuracy. Moreover as we look at normal diffraction pattern images obtained from the experiment we face extreme reduction in resolution as a direct result of brightness saturation, this brightness saturation forces us to use images processed by our network to determine the number of rings.

### 3.3 Ring visualization by network

Since we are using a relatively strong nonlinear material we do not expect to see distinctive rings as we expect from extremely strong nonlinear materials, what we observe are blurry rings that we cannot count with certainty, so instead we introduced our original images to different layers of our network, by doing this we saw that our model doesn't see the diffraction patterns as we do, instead we observed that our model can in fact see and detect 3 distinctive rings as shown in (Fig.5). We can verify this by looking at the same picture in different layers and observe that in all of them we can see 2 rings for low intensities and 3 for more than 168.78 w/cm$^2$ intensity. Note that as the intensity increases our image becomes bigger and we naturally end up with a zoomed in image as we go to into deeper layers.

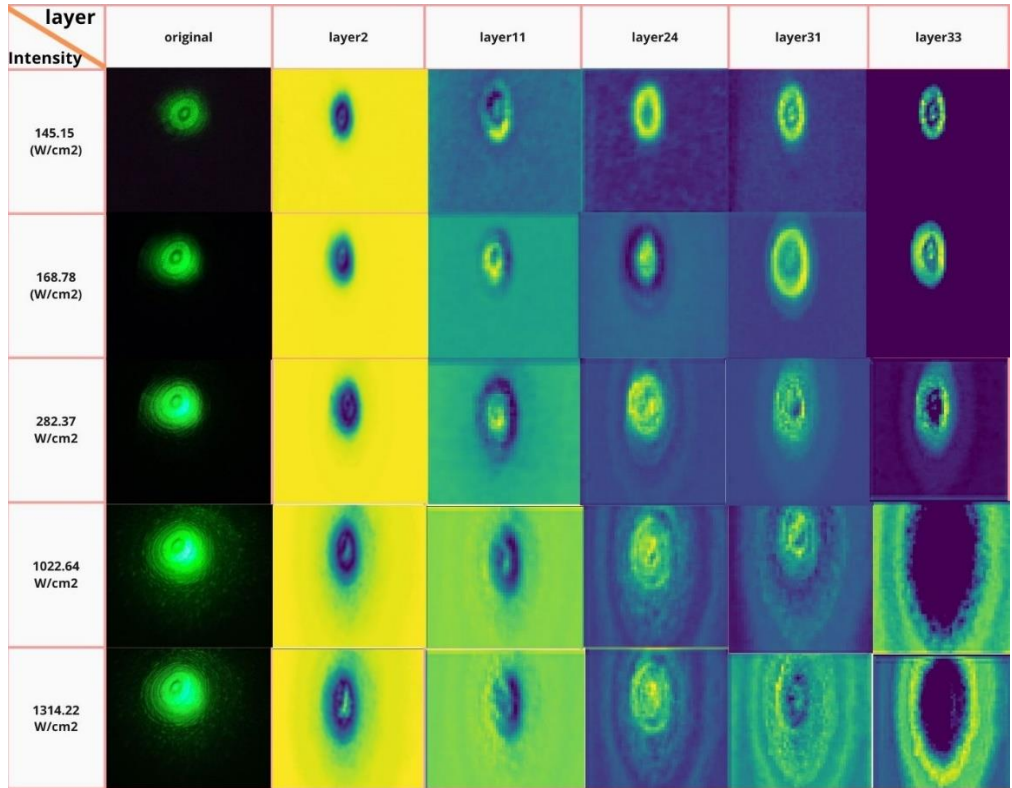

Fig.5 Processed images by the network in different layers

## 4. Conclusion

In this study we fine-tuned a state of the art convolutional neural network architecture to classify the intensity of optical diffraction patterns of a relatively strong nonlinear material and by using phenomena such as optical phase saturation and brightness saturation narrow the true rage of the intensity of an unknown laser beam by mediation.

We were also able to determine the nonlinear refractive index in our relatively strong nonlinear material by visualizing the diffraction patters in the form that our model interprets them. By counting the rings in this form we were able to see and take into account rings that were too blurry to visualize in the original images. No one can deny the power of machine learning and artificial intelligence but these classical networks have computational limitations and cannot simulate or classify increasingly complex systems, hopefully in the near future fault tolerant quantum computers can help us to better expand our knowledge about nature.

### ACKNOWLEDGMENTS

This work was supported by the lasers and optics lab of the department of science and the physics group of the Hakim Sabzevari University.
We also thank Amirhossein Ehsanian for his contributions in the experimental processes for this research.

## AUTHOR DECLARATIONS
**Conflict of Interest**
The authors have no conflicts to disclose.

**Author Contributions**
Behnam Pishnamazi: Data gathering (equal); Formal analysis (equal); Investigation (equal); Software (equal);Code writing; Validation (equal); Writing – original draft (equal). Ehsan Koushki: Formal analysis (equal); Methodology (equal); Resources (equal); Software (equal); Conceptualization (equal); Formal analysis (equal); Writing – original draft (equal); Writing – review & editing (equal) ; Technical Analysis (equal).

## DATA AVAILABILITY
The data that support the findings of this study are available from the corresponding author upon reasonable request.